\documentstyle[preprint,aps,epsfig,psfig]{revtex}

\begin{document}

\draft

\title{QCD Factorization For B Decays To Two Light Pseudoscalars Including
Chirally Enhanced Corrections
\footnote{Supported in part by
National Natural Science Foundation of China and State Commission of
Science and Technology of China}} \vspace{2cm}

\author{ Dongsheng Du${}^{1,2}$, Deshan Yang${}^{2}$ and Guohuai
Zhu${}^{2}$ \footnote{Email: duds@hptc5.ihep.ac.cn,
yangds@hptc5.ihep.ac.cn, zhugh@hptc5.ihep.ac.cn} } \address{${}^1$ CCAST
(World Laboratory), P.O.Box 8730, Beijing 100080, China\\ ${}^2$ Institute
of High Energy Physics, Chinese Academy of Sciences,
 P.O.Box 918(4), Beijing 100039, China
 \footnote{Mailing address}}

\date{\today}

\maketitle

\begin{abstract} 
\indent 

Since b quark mass is not asymptotically large, chirally enhanced
corrections which arise from twist-3 wave functions may be important in B
decays. We thus evaluate the hadronic matrix elements with the emitted
meson described by leading twist and twist-3 distribution
amplitudes $\Phi_p(x)$. After summing over the four "vertex correction"
diagrams, we
obtain the results with infrared finiteness which shows that chirally
enhanced corrections arise from $\Phi_p(x)$ can be consistently included
in QCD factorization. We also briefly discuss the contributions from "hard
spectator" diagrams.
\end{abstract}

\vspace{1.5cm}

{\bf PACS numbers 13.25.Hw 12.38.Bx}

\newpage

\narrowtext 
\tighten

It is well known that two-body, non-leptonic charmless B decays are
crucial for extracting CKM matrix elements. However, due to our ignorance
on how to calculate hadronic matrix elements, we conventionally resort to 
Bjorken's color transparency argument \cite{Bjorken} which lead to "naive
factorization assumption",
\begin{equation}
\langle M_1 M_2 \vert Q \vert B \rangle = \langle M_2 \vert J_1 \vert  
0 \rangle \langle M_1 \vert J_2 \vert B \rangle,
\end{equation}
This assumption makes the hadronic matrix elements scale-independent.
Noting that Wilson Coefficients are scheme- and scale-dependent, the
theoretical
calculations on the branching ratios would then be scheme-
and scale-dependent which
is unacceptable. To save factorization hypothesis, a phenomenological
parameter $N_{eff}$ is introduced \cite{Cheng} which is commonly called 
generalized factorization. However this approach is not satisfactory yet
because in principle $N_{eff}$ is process dependent.

Recently, Beneke, Buchalla, Neubert
and Sachrajda \cite{BBNS1,BBNS3} proposed a 
promising method: in the heavy quark limit, they show that the
emitted light meson can be 
described with leading twist-2 distribution amplitude, the infrared 
divergences of the hard-scattering amplitudes are canceled after summing
over the four "vertex correction" diagrams (Fig.(a)-(d)), which is a
one-loop demonstration 
of Bjorken's color transparency argument \cite{Bjorken}. In the heavy
quark limit, they 
show that the hadronic matrix elements can be expressed as \cite{BBNS1}
\begin{equation}
\langle M_1 M_2 \vert Q \vert B \rangle = \langle M_2 \vert J_1 \vert
0 \rangle \langle M_1 \vert J_2 \vert B \rangle \cdot
[1+\Sigma r_n \alpha_s^n + {\cal{O}}(\Lambda_{QCD}/m_b)].
\end{equation}
If power corrections in $1/m_b$ can be safely neglected, then everything
is perfect.
At the zero order of $\alpha_s$, it would come back to "naive
factorization", and at the higher order of $\alpha_s$, the corrections can
be systematically calculated in Perturbative QCD, which means that the
decay amplitudes
of B meson can be computed from
first principles, and the necessary input
are heavy-to-light form factors and light-cone distribution amplitudes. 
But in the real
world, bottom quark mass is not asymptotically large (but $4.8$ GeV),
and numerically power
suppression may fail in some cases. An obvious and possibly the most
important case
is chirally enhanced power corrections. As pointed out in ref 
\cite{BBNS1},
numerically the enhanced factor 
$r_{\chi}=\frac{2 m_{\pi}^2}{m_b(m_u+m_d)} \simeq 1.18$ which makes
the power suppression completely fail. This parameter is multiplied
by $a_6$ and $a_8$, where $a_6$ is very important numerically in
penguin-dominated B decays. So an evaluation of the hadronic matrix
elements including chirally enhanced corrections may be
phenomenologically
or numerically important. In this letter, we shall
examine this problem in
some detail. 


Chirally enhanced corrections arise from twist-3 light-cone distribution
amplitudes, generally called $\Phi_p(x)$ and $\Phi_{\sigma}(x)$. For light
pseudoscalar mesons, they are defined as \cite{Braun}
\begin{eqnarray}
\langle P(p') \vert {\bar q}(y) {\it i} \gamma_5 q(x) \vert 0 \rangle
&=& f_p \mu_p \int_0^1 {\it du~e}^{i(up' \cdot y + {\bar u}p' \cdot x)}
\phi_p(u), \\
\langle P(p') \vert {\bar q}(y) \sigma_{\mu \nu} \gamma_5 q(x) \vert 0
\rangle &=& if_p \mu_p (p^{\prime}_{\mu} z_{\nu} - p^{\prime}_{\nu}
z_{\mu} ) \int_0^1 {\it du~e}^{i(up' \cdot y + \bar u p' \cdot x)}
\frac{\phi_{\sigma}(u)}{6},
\end{eqnarray}
where $\mu_{p}=\frac{M^2_{p}}{m_u+m_d}$, $z=y-x$. We
notice that in Ref. 
\cite{BBNS1} color transparency is demonstrated in
one-loop level in the heavy
quark limit. If we want to include chirally enhanced
corrections consistently, we should describe the emitted
light meson with leading
twist-2
and twist-3 distribution amplitudes, which means that we should show the
infrared finiteness using twist-3 distribution amplitudes after summing
over the "vertex correction" diagrams. In this paper, we shall restrict
ourselves to $\Phi_p(x)$ while postpone the discussion of
$\Phi_{\sigma}(x)$
to ref \cite{Prepare} because of the complicated derivation in proving the
infrared finiteness of the "vertex correction" diagrams using
$\Phi_{\sigma}(x)$. 

We notice that in
Ref \cite{YY}, the authors have used twist-3 distribution amplitude
$\Phi_p(x)$ to calculate
the strong penguin corrections (Fig.(e)-(f)). The
difference of 
our work from ref \cite{YY} is that we calculate "vertex correction" 
diagrams and show the
infrared finiteness of $a_6$ and $a_8$ at the order of
$\alpha_s$.

 
In the following, we take leading twist and twist-3 wave functions
$\Phi_p(x)$ to
describe the emitted light meson, and we will show the infrared finiteness
under this approach.

The $\vert\Delta B\vert=1$ effective Hamiltonian is given by \cite{Buras}
\begin{equation} 
{\cal{H}}_{eff}= \frac{G_F}{\sqrt{2}}
 \left[ \sum_{q=u,c} v_q \left( C_1(\mu) Q^q_1(\mu)+ C_2(\mu)Q^q_2(\mu)
  + \sum_{k=3}^{10} C_k(\mu)Q_k(\mu)  \right) 
  - v_t(C_{7\gamma}Q_{7\gamma}+C_{8G}Q_{8G})  \right]+h.c., 
\end{equation}
where $v_q=V_{qb}V_{qd}^{*}$(for $b\rightarrow d$ transition) or
 $v_q=V_{qb}V_{qs}^{*}$(for $b\rightarrow s$ transition) and
$C_i(\mu)$ are Wilson coefficients which have been evaluated to
next-to-leading order approximation. The four-quark operators $Q_i$ are 
\begin{equation}
\begin{array}{l}
\begin{array}{ll}
Q^u_1= ( \bar{u}_{\alpha} b_{\alpha} )_{V-A}
         ( \bar{q}_{\beta} u_{\beta} )_{V-A}&
Q^c_1= ( \bar{c}_{\alpha} b_{\alpha} )_{V-A}
         ( \bar{q}_{\beta} c_{\beta} )_{V-A}\\
Q^u_2= ( \bar{u}_{\alpha} u_{\alpha} )_{V-A}
         ( \bar{q}_{\beta} b_{\beta} )_{V-A}&
Q^c_2= ( \bar{c}_{\alpha} c_{\alpha} )_{V-A}
         ( \bar{q}_{\beta} b_{\beta} )_{V-A}\\
Q_3= (\bar{q}_{\alpha} b_{\alpha} )_{V-A}
      \sum\limits_{q'}
     ( \bar{q}^{'}_{\beta} q^{'}_{\beta} )_{V-A}&
Q_4= \sum\limits_{q'} (\bar{q}_{\beta} q^{'}_{\beta} )_{V-A}
     ( \bar{q}^{'}_{\alpha} b_{\alpha} )_{V-A}\\ 
Q_5= (\bar{q}_{\alpha} b_{\alpha} )_{V-A}
      \sum\limits_{q'}
      ( \bar{q}^{'}_{\beta} q^{'}_{\beta} )_{V+A}&   
Q_6= -2\sum\limits_{q'} (\bar{q}_{\beta} q^{'}_{\beta} )_{S+P}
     ( \bar{q}^{'}_{\alpha} b_{\alpha} )_{S-P}\\
Q_7= \frac{3}{2} (\bar{q}_{\alpha} b_{\alpha} )_{V-A}
      \sum\limits_{q'} e_{q'}
     ( \bar{q}^{'}_{\beta} q^{'}_{\beta} )_{V+A}&
Q_8= -3 \sum\limits_{q'} e_{q'} 
    (\bar{q}_{\beta} q^{'}_{\beta} )_{S+P}
    ( \bar{q}^{'}_{\alpha} b_{\alpha} )_{S-P}\\
Q_9= \frac{3}{2} (\bar{q}_{\alpha} b_{\alpha} )_{V-A}
      \sum\limits_{q'} e_{q'}
    ( \bar{q}^{'}_{\beta} q^{'}_{\beta} )_{V-A}&
Q_{10}=\frac{3}{2} \sum\limits_{q'} e_{q'}
       (\bar{q}_{\beta} q^{'}_{\beta} )_{V-A}
     ( \bar{q}^{'}_{\alpha} b_{\alpha} )_{V-A}
\end{array}
\end{array}
\end{equation}
and
\begin{equation}
Q_{7\gamma}=\frac{e}{8\pi^2} m_b \bar{q}_{\alpha} \sigma^{\mu\nu}
(1+\gamma_5) b_{\alpha} F_{\mu\nu}, ~~
Q_{8G}=\frac{g}{8\pi^2} m_b \bar{q}_{\alpha} \sigma^{\mu\nu}
t^{a}_{\alpha \beta} b_{\beta} G^a_{\mu\nu}, ~~(q=d~ {\rm or} ~s).
\end{equation}
The amplitude of the decays of B to two light pseudoscalar mesons
in QCD factorization can be written as:
\begin{equation}
A(B\rightarrow M_1 M_2)=\frac{G_F}{\sqrt{2}}
\sum \limits_{p=u,c} \sum \limits_{i=1,10} v_p a^p_i
\langle M_1 M_2 \vert Q_i \vert B \rangle_F,
\end{equation}
where $v_p$ is CKM factor and
$\langle M_1 M_2 \vert Q_i \vert B \rangle_F$ is the factorized
matrix elements. 

We calculate QCD coefficients $a^p_i$ with the emitted
mesons described by light-cone distribution amplitudes. 
For instance, let we consider the
"vertex correction" diagrams. Twist-3 distribution amplitude $\Phi_p(x)$
makes no contribution when considering $(V-A) \bigotimes (V-A)$ and 
$(S+P) \bigotimes (S-P)$ currents because of their Lorentz structure. As
to $(V-A) \bigotimes (V+A)$ current, there is some subtlety in
regularizing the infrared divergences. If we use dimension
regularization, the infrared finiteness would
not hold for the case of $\Phi_p(x)$ after summing over
those four "vertex correction" diagrams. That is because wave functions
are defined in $4$-dimensions, it may be unconsistent to extend its usage
directly to D-dimensions. Thus we
assign a virtual mass to the gluon propagator and regularize the
infrared integrals in four dimensions. Then the "vertex correction"
contributions of $(V-A) \bigotimes (V+A)$ current to 
$(S+P) \bigotimes (S-P)$ operator is infrared finite:
\begin{equation}
V = \frac{\alpha_s}{4 \pi}\frac{C_F}{N}
 \int_0^1 dx \Phi_p(x) \{
i \pi \log \frac{x}{\bar x} - \frac{1+x}{x}\log \bar x
+\frac{1+\bar x}{\bar x}\log x -Li_2(-\frac{\bar x}{x})
+Li_2(-\frac{x}{\bar x}) \},
\end{equation}
where $\bar x = 1-x$ and $Li_2(x)$ is dilogarithm function.
This means that we can consistently include twist-3 distribution amplitude 
$\Phi_p(x)$ in the framework of QCD factorization. 
The explicit expressions of $a_i^p$ for $i=1$ to $10$ (for symmetric
light-cone distribution amplitudes of light pseudoscalar mesons) are
obtained as:
\begin{eqnarray}
a_1^u&=&C_1+\frac{C_2}{N} + \frac{\alpha_s}{4\pi} \frac{C_F}{N} C_2 F, \\
a_2^u&=&C_2+\frac{C_1}{N} + \frac{\alpha_s}{4\pi} \frac{C_F}{N} C_1 F,\\
a_3&=&C_3+\frac{C_4}{N} + \frac{\alpha_s}{4\pi} \frac{C_F}{N} C_4 F, \\ 
a_4^p&=&C_4+\frac{C_3}{N} -\frac{\alpha_s}{4\pi} \frac{C_F}{N}
\left \{ (\frac{4}{3} C_1 +\frac{44}{3} C_3 + \frac{4f}{3} (C_4+C_6))
\ln \frac{\mu}{m_b} \right .
\nonumber \\
&&+ (G_{M_2}(s_p)-\frac{2}{3})C_1
+(G_{M_2}(0)+G_{M_2}(1)-f_{M_2}^{I} - f_{M_2}^{II}+\frac{50}{3})C_3
\nonumber \\
&&\left . -\frac{2f}{3} C_4
+(3G_{M_2}(0)+G_{M_2}(s_c)+G_{M_2}(1))(C_4+C_6)+G_{M_2,8} C_8
\right \} ,
\\
a_5&=&C_5+\frac{C_6}{N}+\frac{\alpha_s}{4\pi}\frac{C_F}{N} C_6(-F-12),  
\\
a_6^p&=&C_6+\frac{C_5}{N} -\frac{\alpha_s}{4\pi} \frac{C_F}{N}6C_5-
\frac{\alpha_s}{4\pi} \frac{C_F}{N}
\left \{ (C_1+2 C_3+f (C_4+C_6))\ln \frac{\mu}{mb} 
\right . \nonumber \\
&&+(G_{M_2}^{\prime}(s_p)-\frac{7}{12}) C_1 + (G_{M_2}^{\prime}(0) +
G_{M_2}^{\prime}(1)-\frac{7}{6}) C_3 -\frac{7}{12} f C_4 \nonumber \\
&&\left . +(3
G_{M_2}^{\prime}(0)+G_{M_2}^{\prime}(s_c)+G_{M_2}^{\prime}(1))(C_4+C_6)
-\frac{f}{12}C_6 +\frac{3}{2} C_{8G} \right \} ,\\
a_7&=&C_7+\frac{C_8}{N}+\frac{\alpha_s}{4\pi}\frac{C_F}{N} C_8(-F-12),
\\
a_8&=&C_8+\frac{C_7}{N} -\frac{\alpha_s}{4\pi} \frac{C_F}{N} 6 C_7
\nonumber \\
&&+\frac{\alpha_{em}}{9\pi} 
\left \{ -((\frac{C_1}{N}+C_2)+\frac{1}{2}(\frac{C_4}{N}+C_3)
+\frac{1}{2}(\frac{C_6}{N}+C_5))\ln \frac{\mu}{mb} \right . \nonumber \\
&&+(\frac{7}{12}-G_{M_2}^{\prime}(s_p))(\frac{C_1}{N}+C_2)+(\frac{7}{24}-
G_{M_2}^{\prime}(s_c)+\frac{1}{2}G_{M_2}^{\prime}(1))(\frac{C_4}{N}+C_3)
\nonumber \\
&&\left . 
+(\frac{1}{24}-G_{M_2}^{\prime}(s_c)+\frac{1}{2}G_{M_2}^{\prime}(1))
(\frac{C_6}{N}+C_5)-\frac{3}{4} C_{7\gamma} \right \}, \\  
a_9&=&C_9+\frac{C_{10}}{N}+\frac{\alpha_s}{4\pi} \frac{C_F}{N} C_{10} F,
\\
a_{10}^{p}&=&C_{10}+\frac{C_9}{N}+
\frac{\alpha_s}{4\pi} \frac{C_F}{N}C_{9} F \nonumber \\
&&+\frac{\alpha_{em}}{9\pi} \left \{ (-\frac{2}{3} (2(C_2+\frac{C_1}{N})+
(C_3+\frac{C_4}{N})+(C_5+\frac{C_6}{N}))\ln\frac{\mu}{m_b} 
\right . \nonumber \\ 
&&+(\frac{2}{3}-G_{M_2}(s_p))(C_2+\frac{C_1}{N})
+(\frac{1}{3}-G_{M_2}(s_c)+\frac{G_{M_2}(1)}{2})(C_3+\frac{C_4}{N})
\nonumber \\
&&\left . +(-G_{M_2}(s_c)+\frac{G_{M_2}(1)}{2})(C_5+\frac{C_6}{N})
-\frac{1}{2}C_{7\gamma} G_{M_2,8} \right \}.
\end{eqnarray}
Here $N=3$ ($f=5$) is the number of color (flavor),
$C_F=\frac{N^2-1}{2N}$ is the factor of color,
$s_p=m_p^2/m_b^2$ for $p=u,c$ and we define the symbols
in the above expressions as:(most of them are as the same as Beneke's 
except for $G_{M_2}^{\prime}(s)$ and $G^{\prime}(s,x)$)
\begin{eqnarray}
&&F=-12 \ln \frac{\mu}{m_b} -18+f_{M_2}^{I}+f_{M_2}^{II}, \\
&&f_{M_2}^{I}=\int \limits_{0}^{1}~ dx~g(x)\Phi_{M_2}(x),
~G_{M_2,8}=\int \limits_{0}^{1}~ dx~G_8(x) \Phi_{M_2}(x), \\
&&G_{M_2}(s)=\int \limits_{0}^{1}~ dx~G(s,x) \Phi_{M_2}(x), \\
&&G_{M_2}^{\prime}(s)=\int \limits_{0}^{1}~ dx~G^{\prime}(s,x)
\Phi_{M_2}^p(x),
\end{eqnarray}
here $\Phi_{M_2}(x)$($\Phi_{M_2}^p(x)$) is leading twist (twist-3) wave
function of the emitted meson $M_2$, and the hard-scattering functions are
\begin{eqnarray}
&&g(x)=3 \frac{1-2x}{1-x} \ln x - 3 i \pi, ~~G_8(x)=\frac{2}{1-x}, \\  
&&G(s,x)=-4 \int \limits_{0}^{1}~ du~u(1-u) \ln (s-u(1-u)(1-x)-i
\epsilon), \\
&&G^{\prime}(s,x)=\frac{3}{4}G(s,x).
\end{eqnarray}
As to $f_{M_2}^{II}$ which labels the contributions from the hard
spectator scattering diagrams (Fig.(g)-(h)), 
We take the wave function of B meson
as $\gamma_5 (\slash{\hskip -2.5mm}P_B -M_B )\Phi_B
(\xi)$ and find that,
when considering twist-3 $\Phi_p(x)$ distribution contributions, the hard
spectator
scattering contributions are proportional to : \\
(i) for the case of $(V-A) \otimes (V-A)$ and 
$(S+P) \otimes (S-P)~$currents ,
\begin{equation}
 \int \limits_0^1 ~ d \xi~ \frac{\Phi_B( \xi )}{\xi}
\int \limits_0^1 ~ dy~ \frac{\Phi_{M_1}(y)}{y}
\int \limits_0^1 ~ dx~ \left \{ \frac{\Phi_{M_2}(x)}{x}+
\frac{\Phi^p_{M_2}(x)\frac{\mu_{M_2}}{M_B}(x-(1-x))}{x(1-x)}
\right \};
\end{equation}
(ii) for the case of $(V-A) \otimes (V+A)$ currents,
\begin{equation}
 \int \limits_0^1 ~ d \xi~ \frac{\Phi_B( \xi )}{\xi}
\int \limits_0^1 ~ dy~ \frac{\mu_{M_1}}{M_B} \frac{\Phi^p_{M_1}(y)}{y}
\int \limits_0^1 ~ dx~
\frac{\Phi^p_{M_2}(x)\frac{\mu_{M_2}}{M_B}(x-(1-x))}{x(1-x)}.
\end{equation}
Under the symmetric distributions of final state light mesons, the
logarithmically
divergent integrals are canceled after summing over two hard spectator
scattering diagrams.
As a result there is no hard spectator scattering contributions to
$a_6$ and
$a_8$, the contribution of hard spectator scattering to other $a_i^p$ is
as the same as 
ref \cite{BBNS1,our}:
\begin{equation}
f_{M_2}^{II}=\frac{4 \pi^2}{N}
\frac {f_{M_1}f_B}{F^{B\rightarrow M_1}_{+}(0) m_B^2}
\int \limits_{0}^{1}~ d\xi~ \frac{\Phi_B(\xi)}{\xi}
\int \limits_{0}^{1}~ dx~ \frac{\Phi_{M_1}(x)}{x} \int \limits_{0}^{1}~
dy~ \frac{\Phi_{M_2}(y) }{y}.
\end{equation}
In ref \cite{BBNS2}, the authors have discussed the contributions of
asymmetric
distributions and find that numerically this effect is very small. In
our case, when considering the asymmetric distributions, there would
appear divergent integrals, but in this case the asymmetric distribution 
corrections would
also be small if we parametrized the divergent integrals as an
unknown parameter(as what have done in ref \cite{BBNS2}) and
could be safely neglected.

We notice that the above approach of evaluating "hard spectator"
contribution
is naive. For instance, the scale of "hard spectator" contribution should
be different from the "vertex correction" contribution. While it seems 
reasonable, for the "vertex correction" diagrams, to take the
scale
$\mu \sim {\cal O}(m_b)$ to avoid large logarithm 
$\alpha_s \log \frac{\mu}{m_b}$, a natural choice  of the scale of "hard
spectator" contribution may be around ${\cal O}(1~GeV)$ because the
average 
momentum square of the exchanged gluon is about $1~ GeV^2$. 
Another disturbing feature of "hard spectator" contribution is that, as
have been pointed out in
ref \cite{BBNS2}, when including the contribution of $\Phi_{\sigma}$,
there would appear divergent integral $\int_0^1 dx \frac{1}{x}$ even if
the symmetric distribution amplitude is applied. This divergent integral
implies
that the dominant contribution comes from the end-point region, or in
another word, it is dominated by soft gluon exchange. However the
transverse momentum may not be omitted in the end-point region, if so, the
corresponding divergent integral would then changed to:
\begin{equation}
\int dx d^2 k_T \frac{\phi(x,k_T)}{x \xi m_b^2 +k^2_T}.
\end{equation}
As an illustration, we do not consider the $k_T$ dependence of wave
functions (though it is certainly not a good approximation), then the
above integral proportions to:
\begin{equation}
\int \frac{dx d k^2_T}{x \xi m_b^2 +k^2_T} \propto
\int \frac{dx dy}{x+y}.
\end{equation}
The above integration is convergence now, furthermore it is not dominated
by end-point contribution. This illustrates that the
treatment of "hard spectator" diagrams may need further
discussions. 
 


In summary, we consider some chirally enhanced corrections arise from
twist-3 distribution amplitudes $\Phi_p(x)$. To include
chirally enhanced corrections consistently in QCD
factorization,  
we describe the emitted light meson with leading twist and twist-3
$\Phi_p(x)$ distribution amplitudes and show the infrared finiteness of
"vertex correction" diagrams
(fig.(a)-(d)). We also briefly discuss the disturbing "hard spectator" 
contributions.
\section*{Acknowledgements}
This work is supported in part by National Natural
Science Foundation of China and State Commission of
Science and Technology of China.
\narrowtext
\tighten


\begin{table}
\vspace*{1cm}
\begin{tabular}{ccccc}
QCD &  \multicolumn{2}{c} {$\mu=5.0~GeV$} &
\multicolumn{2}{c} {$\mu=2.5~GeV$} \\  
Coefficients & NLO & LO  & NLO & LO \\ \hline
$a_1^u$ & $1.043+0.012 i$   & $1.019$  &
          $1.067+0.025 i$   & $1.039$  \\
$a_2^u$ & $0.034-0.075 i$ & $0.184$  &
          $-0.011-0.104 i$ & $0.104$  \\ \hline
$a_3$   & $0.007+0.002 i$ & $0.003$  &
          $0.010+0.004  i$ & $0.005$   \\
$a_4^u$ & $-0.030-0.014 i$ & $-0.029$ &
          $-0.033-0.018 i$ & $-0.041$ \\   
$a_4^c$ & $-0.035-0.006 i$ & $-0.029$ &
          $-0.040-0.006 i$ & $-0.041$  \\
$a_5$   & $-0.007-0.003 i$ & $-0.005$ & 
          $-0.009-0.006 i$ & $-0.010$ \\ 
$r_{\chi} a_6^u$ & $-0.047-0.003 i$ & $-0.044$ &
          $-0.050-0.003 i$ & $-0.052$ \\
$r_{\chi} a_6^c$ & $-0.049-0.006 i$ & $-0.044$   &
          $-0.053-0.006 i$ & $-0.052$ \\ \hline
$a_7 \times 10^{5}$   & $11.1+2.7 i$     & $9.1$   &
          $2.8+5.5 i$     & $3.6$  \\    
$r_{\chi} a_8^u \times 10^{5}$ & $49.0-3.0 i$     & $45.3$   &
          $57.0-1.2 i$     & $49.2$ \\   
$r_{\chi} a_8^c \times 10^{5}$ & $47.3-4.6 i$     & $45.3$   &
          $56.3-1.9 i$     & $49.2$ \\   
$a_9 \times 10^{5}$   & $-939.8-13.3 i$  & $-913.3$ &
          $-969.6-25.4 i$  & $-941.4$ \\
$a_{10}^u \times 10^{5}$ & $22.4+57.8 i$   & $-135.7$ &
           $68.1+88.9 i$   & $-68.0$ \\
$a_{10}^c \times 10^{5} $ & $19.1+63.0 i$   & $-135.7$ &
           $66.2+91.9 i$   & $-68.0$ \\   
\end{tabular}
          
\vspace{0.5cm}
\caption{ The QCD coefficients $a_i^p(\pi \pi)$ at NLO and LO for the
renormalization scales at $\mu=5~ GeV$ and $\mu=2.5~GeV$ } 

\end{table}


\begin{figure}[tb]
\vspace*{1cm}
\centerline{\epsfig{figure=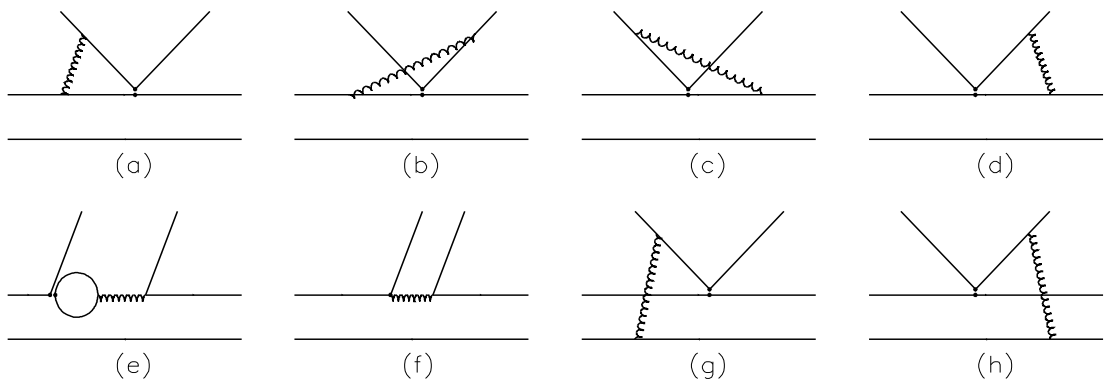,height=6cm,width=15cm,angle=0}}
\vspace*{1.cm}
\caption{Order of $\alpha_s$ corrections to hard-scattering kernels. 
The upward quark lines represent the ejected quark pairs
from b quark weak decays.}
\end{figure}


\end{document}